# Electrostatics of Pagels-Tomboulis effective model

by

## H. Arodź, M. Ślusarczyk and A. Wereszczyński

Marian Smoluchowski Institute of Physics,
Jagellonian University, Reymonta 4, Kraków, Poland

### Abstract

Long time ago Pagels and Tomboulis have proposed a model for the nonperturbative gluodynamics which in the Abelian sector can be reduced to a strongly nonlinear electrodynamics. In the present paper we investigate Abelian, static solutions with external charges in that model. Nonzero total charge implies that the corresponding field has infinite energy due to slow fall off at large distances. For a pair of opposite charges the energy is finite – it grows like $R^\alpha$, $0 < \alpha < 1$ with the distance $R$ between the charges.

# 1  Introduction

Nontrivial features of QCD such as the asymptotic freedom or the quark confinement are due to its gluonic part. Unfortunately, in spite of numerous efforts dynamics of non-Abelian gauge fields in a nonperturbative regime is not fully understood yet. Popular method of investigating it is the lattice approximation. It is the most successful one up to now, yet it has well-known drawbacks. There have been several attempts to apply the approach which has turned out to be very fruitful in condensed matter physics, that is to construct Ginzburg-Landau type effective models which would capture the main features of the nonperturbative gluonic dynamics already on the classical level. In the condensed matter physics, precisely the Ginzburg-Landau models are the main tool for investigating such nonperturbative phenomena as, for example, vortices in superfluids and in superconductors, or hedgehogs in nematic liquid crystals.

In the literature there are several proposals for effective models for the nonperturbative gluonic dynamics. The most popular ones are the dual superconductor model [1,2], the colour dielectric model [3,4], and the stochastic vacuum model [5,6]. Each of them has its own problems. It seems that the satisfactory effective model for the nonperturbative gluonic sector has not been found yet, but this does not mean that such model necessarily will be very different from the ones already considered in the literature.

The effective Ginzburg-Landau model is expected to give a good description of physics already on the classical level (a mean field approximation). When constructing such a model, the first step consists in choosing dynamical variables, that is the field content of the model. In the case of gluodynamics almost all models known to us contain Yang-Mills fields or at least a $U(1)$ field if one assumes a kind of Abelian projection, and sometimes other fields. Then, it is clear that the effective model for the nonperturbative gluonic dynamics has to be drastically different from typical field theoretical models. First, it should not allow for existence of massless, freely propagating vector particles (gluons). Second, it should predict absence of coloured states, and a confining force between (anti)quarks in colour neutral states. Third, it should predict existence of massive glueballs which could propagate freely. Finally, the model should explicitly contain a mass scale – for the massive glueballs as well as in order to reproduce the trace anomaly of energy-momentum tensor. It is a well-known fact that classical Yang-Mills theory with the standard Lagrangian obeys none of these requirements.

In the present paper we investigate a model which has been proposed long time ago by Pagels and Tomboulis [7]. In a sense it can be regarded as a version of the colour dielectric model without a dilaton type scalar field - the pertinent Lagrangian is a function of Yang-Mills field strength only. The model is too simple in order to provide a very good approximation to the real-life nonperturbative gluodynamics in QCD, but it looks rather promising as far as the four requirements specified above are concerned. Our general goal is to study such highly nonstandard field theoretical models in order to clarify their physical contents.

Pagels and Tomboulis calculated electric field of a single point source in their model. They found that in comparison with the standard Coulomb field it is less singular close to the charge and much stronger far away from it. The corresponding energy is infinite due to the large distance behaviour rather than the singularity at the origin. This result implies a confinement of charges. However, one should also check that a pair of charges $Q$, $-Q$ has finite energy, otherwise also the quark-antiquark pairs would disappear from the physical spectrum and the model would be wrong. Because of nonlinearity of Gauss law in the model, one can not compute the electric field of the pair by a simple superposition. In the present paper, using a combination of analytic and numerical methods, we check that the energy of the pair $Q$, $-Q$ is finite if a parameter $\delta > 1/4$. We also calculate distribution of electric field around the charges. It turns out to be very similar to a flux-tube, that is it is localised around the straight line connecting the charges, but an effective string tension depends on the distance $R$ between the charges. The total energy behaves like $R^\alpha$, where $\alpha = (4\delta - 1)/(4\delta + 1)$ and $\delta$ is a parameter. Thus, the charges are confined if $\delta > 1/4$ because then $0 < \alpha < 1$. The effective string tension behaves like $R^{\alpha-1}$. For $\delta = 1/4$ we obtain the logarithmic behaviour of the energy.

Interestingly enough, for $\delta = 3/4$ we obtain the $\sqrt{R}$ behaviour of the energy, which is in agreement with a phenomenological potential found in fits to spectra of heavy quarkonia [8, 9].

Our paper is organised as follows. In Section 2 we recall the model and we derive field equations. We also point out that in that model classical vacuum is strongly degenerate. In Section 3 we calculate the energy of the system of two static, opposite charges. Several remarks about the model are collected in Section 4.

## 2 The Pagels-Tomboulis model

Let us denote by $\mathcal{F}_2$ the standard Yang-Mills invariant[1]

$$\mathcal{F}_2 = \frac{1}{2} F^a_{\mu\nu} F^{a\mu\nu} = B^{ai} B^{ai} - E^{ai} E^{ai}, \tag{1}$$

---

[1]We choose the metric in Minkowski space-time with the signature $(+1, -1, -1, -1)$.

where the field strength $F^a_{\mu\nu}$ has the usual form

$$F^a_{\mu\nu} = \partial_\mu A^a_\nu - \partial_\nu A^a_\mu - g f_{abc} A^b_\mu A^c_\nu \qquad (2)$$

and $E^{ai} = F^a_{0i}$, and $B^{ai} = -\frac{1}{2}\varepsilon_{ikl}F^a_{kl}$ are the non-Abelian electric field and magnetic induction field, correspondingly. We will consider only $SU(2)$ gauge group, hence $a, b, c = 1, 2, 3$. The model we would like to investigate has Lagrangian $\mathcal{L}$ of the form $\mathcal{L} = \ell(\mathcal{F}_2)$, where $\ell$ is a nontrivial function which we shall choose shortly. In a connection with QCD models of this kind have been proposed long time ago [7], [10], [11]. The Abelian Born-Infeld electrodynamics is even older [12], but its aim is to modify only physics of strong fields, while in the QCD case the main problem is with weak fields. The same remark applies to the quite popular recently non-Abelian Born-Infeld actions [13].

Pagels and Tomboulis [7] considered Lagrangians of the form

$$L_{eff}(A_\mu) = \frac{1}{2}\frac{F_2}{\bar{g}^2(t)} \qquad (3)$$

in Euclidean space-time. In formula (3) $F_2 = F^a_{\mu\nu}F^a_{\mu\nu}/2$ is non-negative due to the Euclidean metric, $t = \ln\frac{F_2}{\mu^4}$, and $\bar{g}(t)$ is the effective coupling constant determined from the equation

$$t = \int_{g_0}^{\bar{g}(t)} \frac{dg}{\beta(g)}, \qquad (4)$$

where $\beta(g)$ is the Gell-Mann, Low function. The Lagrangian (3) has been proposed in [7] as an Ansatz consistent with renormalization group invariance of the effective action $\int d^4 x L_{eff}$, and also with the trace anomaly. The perturbative 1-loop result for the $\beta(g)$ function:

$$\beta(g) \cong -b_0 g^3, \quad b_0 > 0, \qquad (5)$$

where $b_0$ is a positive constant, gives

$$\frac{1}{\bar{g}^2(t)} = \frac{1}{g_0^2} + 2b_0 t, \qquad (6)$$

and in consequence

$$L_{eff} = \frac{1}{2}\frac{F_2}{g_0^2} + b_0 F_2 \ln\frac{F_2}{\mu^4}. \qquad (7)$$

Here $1/g^2$ is the integration constant, equal to the value of the coupling constant at the subtraction point $\mu$. Another case, also considered in [7], corresponds to

$$\beta(g) = -\delta g, \qquad (8)$$

where $\delta > 0$ is a constant. Now Eq. (4) gives

$$\frac{1}{\bar{g}^2(t)} = \frac{1}{g_0^2}\left(\frac{F_2}{\mu^4}\right)^{2\delta} \tag{9}$$

and

$$L_{eff} = \frac{1}{2g_0^2}F_2\left(\frac{F_2}{\mu^4}\right)^{2\delta}. \tag{10}$$

Notice that

$$\left(\frac{F_2}{\mu^4}\right)^{2\delta} = \exp\left[2\delta \ln\left(\frac{F_2}{\mu^4}\right)\right] = 1 + 2\delta \ln\frac{F_2}{\mu^4} + \ldots, \tag{11}$$

hence Lagrangian (10) with $\delta = b_0 g^2$ reduces to (7) when $2\delta \ln\frac{F_2}{\mu^4} \ll 1$, that is when $F_2$ is close to the subtraction point $\mu$. One may regard Lagrangian (10) as a resumation of powers of $\ln(F_2/\mu^4)$ in the effective action functional $\Gamma(A_\mu)$ for the Yang-Mills field. From this viewpoint, $\delta$ can be associated with the anomalous dimension of the $F_2$ operator.

The corresponding effective models in Minkowski space-time are obtained by replacing the Euclidean $F_2$ by $\mathcal{F}_2$. In order to avoid problems with the sign of $\mathcal{F}_2$ and complex-valued Lagrangians we write $\ln \frac{F_2}{\mu^4}$ in (7) as $\frac{1}{2}\ln\left(\frac{F_2}{\mu^4}\right)^2$, and $\left(\frac{F_2}{\mu^4}\right)^{2\delta}$ as $\left(\frac{F_2^2}{\mu^8}\right)^\delta$ in (10).

The model corresponding to $L_{eff}$ (7) was considered by Adler and Piran [11] in the case where only non-Abelian electric fields $F_{0i}^a$ were present - it gave a confining force between two point-like opposite charges, at least in Abelian sector obtained by assuming that $A_\mu^a(x) = \delta_3^a A_\mu(x)$. However, it turns out that if one allows for the presence of magnetic fields, the corresponding energy density is not bounded from below. Moreover, Lagrangian (7) contains the standard kinetic term $(\partial_\mu A_\nu^a - \partial_\nu A_\mu^a)^2$ hence the gluons could propagate freely. For these reasons, we conclude that the logarithmic model obtained from (7) is not satisfactory.

Let us turn to the model with $L_{eff}$ given by formula (10). When passing to Minkowski space-time we change notation a little bit. In Minkowski space-time we take Lagrangian of the form

$$\mathcal{L} = -\frac{1}{2}\mathcal{F}_2\left(\frac{\mathcal{F}_2^2}{\Lambda^8}\right)^\delta, \tag{12}$$

where $\delta > 0$, instead of Lagrangian (10). The initial coupling constant $g_0$ has been included into $\Lambda$ which is regarded as an empirically fixed energy scale in the model.

Components of energy-momentum tensor corresponding to Lagrangian (12) have the form

$$T_{00} = \frac{1}{2}\left[\vec{B}^2 + (1+4\delta)\vec{E}^2\right]\left(\frac{\mathcal{F}_2^2}{\Lambda^8}\right)^\delta, \tag{13}$$

$$T_{0i} = -(1+2\delta)\varepsilon_{iks}E^{ak}B^{as}\left(\frac{\mathcal{F}_2^2}{\Lambda^8}\right)^\delta \tag{14}$$

and

$$T_{ik} = \left\{\frac{1}{2}\left[\vec{E}^2 + (1+4\delta)\vec{B}^2\right]\delta_{ik} - (1+2\delta)\left[E^{ai}E^{ak} + B^{ai}B^{ak}\right]\right\}\left(\frac{\mathcal{F}_2^2}{\Lambda^8}\right)^\delta. \tag{15}$$

We see that in this model $T_{00} \geq 0$.

Lagrangian (12) does not contain the standard kinetic term for the gauge fields. Therefore, it is by no means clear what are physical, propagating excitations in the model. Let us introduce the non-Abelian dielectric induction

$$D^{ai} = \frac{\partial \mathcal{L}}{\partial E^{ai}} = (1+2\delta)E^{ai}\left(\frac{\mathcal{F}_2^2}{\Lambda^8}\right)^\delta \tag{16}$$

and the non-Abelian magnetic field:

$$H^{ai} = -\frac{\partial \mathcal{L}}{\partial B^{ai}} = (1+2\delta)B^{ai}\left(\frac{\mathcal{F}_2^2}{\Lambda^8}\right)^\delta. \tag{17}$$

The modified Yang-Mills equations following from Lagrangian (12) can be written in the form

$$\nabla^a_{ib}D^{bi} = 0, \quad \nabla^a_{0b}D^{bk} - \varepsilon_{kir}\nabla^a_{ib}H^{br} = 0, \tag{18}$$

where

$$\nabla^a_{\mu b} = \delta^a_b \partial_\mu + f_{bca}A^c_\mu$$

is the covariant derivative. Of course in addition to these equations we have the standard non-Abelian Bianchi identities for $E^{ai}$ and $B^{ai}$:

$$\nabla^a_{ib}B^{bi} = 0, \quad \nabla^a_{0b}B^{bi} + \varepsilon_{ijk}\nabla^a_{jb}E^{bk} = 0. \tag{19}$$

These identities are equivalent to formula (2).

It is clear that any fields such that

$$B^{ai}B^{ai} = E^{ai}E^{ai} \tag{20}$$

obey the field equations because then $\mathcal{F}_2 = 0$. However, they have vanishing energy-momentum tensor $T_{\mu\nu}$ and therefore they should be regarded as vacuum fields! The model has enormously degenerate classical vacuum state, which includes, in particular, plane waves. In the standard Yang-Mills theory, small amplitude classical plane waves correspond to the perturbative gluons. It is encouraging that they do not belong to the spectrum of physical excitations of Pagels-Tomboulis model. We plan to address the issue of physical excitations in Pagels-Tomboulis model in a forthcoming paper [14]. In the present paper we show that the model can describe the confinement of quarks.

Formula (16) implies that the dielectric induction is smaller than the electric field if $\frac{\mathcal{F}_2^2}{\Lambda^8} < (1 + 2\delta)^{-1/\delta}$, that is the model implies antiscreening for weak fields. For strong fields we have the usual screening.

## 3 Static solutions with external point charges

**a) The case of non-vanishing total charge**
Equations (18), (19) are even more complicated than the standard Yang-Mills equations, which are obtained for $\delta = 0$. The fact that the classical Yang-Mills theory is not confining can be seen already from an analysis of its Abelian sector. For this reason, we check the Abelian sector of Pagels-Tomboulis model ($\delta > 0$). Let us remind that that sector is constituted by gauge potentials $A_\mu^a$ with only one colour component, identical for all $\mu = 0, 1, 2, 3$. For example, we may take

$$A_\mu^a = \delta_3^a A_\mu(x). \tag{21}$$

In the Abelian sector, the covariant derivatives in Eqs. (18), (19) reduce to the ordinary ones and we may omit the superscript $^a$. Nevertheless, the equations remain nonlinear if $\delta > 0$ – the cases $\delta = 0$ and $\delta > 0$ are drastically different. In particular, if $\delta \geq 1/4$ the Abelian sector of Pagels-Tomboulis model is compatible with the confinement of charges.

Let us consider first a smooth, spherically symmetric distribution $j_0(r)$ of an external, static charge in a finite region around the origin. The Gauss law has the form

$$\vec{\nabla}\vec{D} = j_0(r), \tag{22}$$

and the electric field obeys the condition

$$\nabla \times \vec{E} = 0, \tag{23}$$

with $\vec{D}$ and $\vec{E}$ related by formula (16) (we omit the superscript $a = 3$). Because of the nonlinearity, solutions of Eqs. (22), (23) can not be obtained by superposition of solutions for point charges. Nevertheless, spherically symmetric solutions follow easily from Gauss law (22). Far away from the charge

$$\vec{D} = \frac{Q}{4\pi} \frac{\vec{r}}{r^3}, \quad \frac{\vec{E}}{\Lambda^2} = sign(Q)(1+2\delta)^{-1/(1+4\delta)} \left(\frac{|Q|}{4\pi \Lambda^2 r^2}\right)^{\frac{1}{1+4\delta}} \frac{\vec{r}}{r}, \quad (24)$$

where

$$Q = \int d^3 r j_0(r).$$

Thus, both fields vanish when $r \to \infty$, namely $|\vec{D}| \sim r^{-2}$, $|\vec{E}| \sim r^{-2/(1+4\delta)}$. Simple calculation shows that the total energy of the gauge field, $\int d^3 r T_{00}$, is infinite if $\delta \geq 1/4$, due to the behaviour of $\vec{E}$ at large $r$. This fact is not changed by inclusion of the interaction energy of the external charge with the gauge field, $\int d^3 r j_0 A_0$, because this integral is finite. For comparison, in the Yang-Mills case the energy of the gauge field (24) (now with $\delta = 0$) is finite.

This result suggests that in Pagels-Tomboulis model states with nonvanishing total charge $Q$ are not physically feasible. In order to turn this suggestion into a theorem one would have to find solutions of the non-Abelian gauss law

$$\nabla^a_{ib} D^{bi} = \delta^a_3 j_0(r),$$

together with the remaining equations (18), (19), without the simplifying assumption (21). Even in the Yang-Mills case this is a highly nontrivial task, see e. g. [15] for a review. In particular, Kiskis [16] showed that in the Yang-Mills case exist solutions with arbitrarily small (but positive) energy. Such solutions in general are not static and have nonvanishing magnetic fields. However, his reasoning is based on linearity of Yang-Mills equations in the Abelian sector, and therefore it can not be repeated in Pagels-Tomboulis model. This is encouraging, nevertheless the question whether finite energy solutions with $Q \neq 0$ are completely excluded remains open.

**b) Dipol-like external charge density**

Effective model for QCD should allow for finite energy colour singlet quark-antiquark states [2]. Therefore, we shall now consider a dipole-like external charge.

The results of subsection 3a) are valid also when the external charge is point-like, $j_0(r) = Q\delta(\vec{r})$. Then, the resulting energy density has a singularity at $r = 0$ which is integrable if $\delta > 1/4$. Therefore, it is not important for our purpose whether the two charges forming the dipole are point-like or spatially extended. For simplicity, we assume that they are point-like. Thus, we now take

$$j_0 = q\delta(x)\delta(y)\left[\delta(z + R/2) - \delta(z - R/2)\right] \qquad (25)$$

where $q > 0$, and again we look for solutions of Eqs. (22), (23) in the Abelian sector.

---
[2]In the case of SU(3) gauge group colour singlets built of three quarks are related to the quark-antiquark states because a diquark can be regarded as an antiquark as far as colour charge is concerned.

We would like to check that the corresponding total energy $\mathcal{E}$ of the gauge field, given by the integral $\mathcal{E} = \int d^3 r T_{00}$, is finite. First, simple dimensional analysis shows that

$$\mathcal{E} = c_0 |q|^{\frac{2+4\delta}{1+4\delta}} \Lambda^{\frac{8\delta}{1+4\delta}} R^{\frac{4\delta-1}{4\delta+1}}, \qquad (26)$$

where $c_0$ is a numerical constant. This formula follows from the fact that

$$\mathcal{E} \sim |q|^{\frac{2+4\delta}{1+4\delta}} \Lambda^{\frac{8\delta}{1+4\delta}},$$

as implied by Eq. (22) and formulas (13), (16). The exponent of $R$ is dictated by the requirement that $\mathcal{E}$ has the dimension $cm^{-1}$. Thus, it remains to show that the constant $c_0$ is finite. Because singularities at the point-like charges are of integrable type, finiteness of $c_0$ depends solely on behaviour of the fields at the spatial infinity.

Following Adler and Piran, [11], [17], [18], we express the dielectric induction $\vec{D}$ by the dual potential $\vec{C}$:

$$\vec{D} = \nabla \times \vec{C}. \qquad (27)$$

Then, at all points where $\vec{C}$ is sufficiently smooth, automatically $\nabla \vec{D} = 0$. However, the presence of source (25) on the r.h.s. of Eq. (22) implies that $\vec{C}$ can not be regular everywhere – it has to have a singularity akin to the Dirac string for magnetic monopoles. In our case it is natural to assume that the string connects the two point charges. To exploit the axial symmetry of the problem it is natural to introduce the cylindrical coordinates $(\rho, \phi, z)$ instead of the Cartesian ones. The coordinates of the point sources are then $\rho = 0$, $z = \pm R/2$. As in [11], [17], [18] we assume that the dual potential $\vec{C}$ can be expressed by a scalar flux function $\Phi(\rho, z)$, which is defined as follows

$$\vec{C} = \frac{\hat{\phi}}{2\pi\rho}\Phi, \qquad (28)$$

where $\hat{\phi}$ is the unit vector tangent to the $\phi$ coordinate line.

Equation (23) may be rewritten as

$$\nabla \times \left(\frac{\vec{D}}{\varepsilon}\right) = 0, \qquad (29)$$

where

$$\varepsilon = (1 + 2\delta)\left(\frac{\vec{E}^2}{\Lambda^4}\right)^{2\delta} \qquad (30)$$

is the dielectric function. The Ansatz (28) reduces Eq.(29) to

$$\nabla(\sigma\nabla\Phi) = 0, \qquad (31)$$

where

$$\sigma = \left(\frac{1}{\rho}\right)^{\frac{2+4\delta}{1+4\delta}} \left(\frac{1}{|\nabla\Phi|^2}\right)^{\frac{2\delta}{1+4\delta}}. \qquad (32)$$

It remains to fix boundary conditions for $\Phi$. Formulas (27), (28) imply that

$$\begin{aligned}\Phi &= 0 \quad \text{for} \quad \rho = 0, \ |z| > R/2, \\ \Phi &= q \quad \text{for} \quad \rho = 0, \ |z| < R/2.\end{aligned} \qquad (33)$$

To obtain, for example, the second line in (33), consider a sphere surrounding one of the charges with a small hole around the point at which the sphere is punched by the segment of the z-axis connecting the charges. The flux of the $\vec{D}$ field through such surface is equal to line integral of $\vec{C}$ over the boundary of the hole. Shrinking the hole to the point of the intersection we find that the value of $\Phi$ at that point is equal to the total flux $Q$ of $\vec{D}$ through the sphere.

We also assume that

$$\Phi \to 0 \quad \text{for} \quad \rho^2 + z^2 \to \infty. \qquad (34)$$

This condition is justified by the expectation that angular dependence of $\Phi$ at the spatial infinity would increase the total energy. Because it has been assumed that $\Phi$ vanishes on the z-axis if $|z| > R/2$, it has to vanish in all other directions.

To summarize, the problem of solving the set of field equations for the electric field $\vec{E}$ with given charge density (25) in three spatial dimensions is reduced to one sourceless equation (31) in the region $\rho > 0, z \in (\infty, -\infty)$, which can be regarded as a cylindrical shell with the outer radius going to $\infty$ and the inner one to 0. The boundary conditions are given by (33), (34).

Equation (31) can be rederived from the condition that the flux function minimizes the total field energy $\mathcal{E} = \int d^3 r T_{00}$ under the boundary conditions (33), (34). Formulas (13), (27) and (28) give

$$T_{00} = \frac{1}{2}(1+4\delta)\left[2\pi(1+4\delta)\rho\right]^{\frac{-2-4\delta}{1+4\delta}} \Lambda^{\frac{8\delta}{1+4\delta}} |\nabla\Phi|^{\frac{2+4\delta}{1+4\delta}}, \tag{35}$$

where $|\nabla\Phi| = \sqrt{(\nabla\Phi)^2}$. One can easily construct examples of the flux function which obey the boundary conditions and have finite total energy. For example, we may take

$$\Phi = \frac{q}{2}\left(\frac{z+R/2}{\sqrt{\rho^2+(z+R/2)^2}} - \frac{z-R/2}{\sqrt{\rho^2+(z-R/2)^2}}\right). \tag{36}$$

This function has been obtained from the sum of the dual potentials for two Dirac monopoles of opposite charges, and it implies that the Dirac string just connects the charges $q, -q$ along the z-axis. The function (36) obeys the boundary conditions, but it does not obey Eq. (31) unless $\delta = 0$. Nevertheless, the finite value of $\mathcal{E}$ corresponding to it may be taken as an upper bound for the total field energy of the charges $q, -q$.

Solutions of Eq.(31) can be found with the help of numerical computations. We use the standard iterative procedure [17]. Due to the symmetry of the problem it is sufficient to restrict our approach to the region $z \geq 0$, $\rho \geq 0$. The continuous variables $\rho, z$ are replaced by the computational lattice with $(n_\rho + 1) \times (n_z + 1)$ sites; the flux function $\Phi$ reduces to the discrete set of values on the lattice $\Phi_{i,j}$ where $i = 0\ldots n_\rho$ and $j = 0\ldots n_z$. The point charge $q$ is put on the site of the computational lattice with $i = 0, j = n_q$ where $0 < n_q < n_z$. The boundary conditions (33), (34) are replaced by:

$$\Phi_{0,j} = q, \quad 0 \leq j < n_q, \tag{37}$$
$$\Phi_{0,n_q} = q/2,$$
$$\Phi_{0,j} = 0, \quad n_q < j \leq n_z,$$
$$\Phi_{n_\rho,j} = 0, \quad 0 \leq j \leq n_z,$$
$$\Phi_{i,n_z} = 0, \quad 0 \leq i \leq n_\rho.$$

In the first step, sites of the computational lattice are populated by arbitrary values. Next, new $\Phi_{i,j}$ are computed using a discretized version of the field equation (31). The iterative procedure stops when the difference between actual value of $\Phi_{i,j}$ and the one computed in the previous step is less then a chosen accuracy. By assumption, such approximate solution obeys the homogeneous boundary condition (34) already at finite $\rho$ and $z$.

In our computations we used $900 \times 900$ mesh. Some sample results were also obtained for bigger meshes but the results changed insignificantly. We have repeated our procedure for some values of the charge separation distance $R$ and fixed size of the lattice. The final results were obtained for $\delta = 0.75$, and $q = 1.1$. The results are presented in Figs. 1 – 5. Fig. 1 presents the flux function $\Phi$, Fig. 2 the profile of $\Phi$ for $z = 0$. From the definition of $\Phi$ the dielectric induction $\vec{D}$ as well as the energy density $\varepsilon$ may be derived. The rescaled (dimensionless) energy density $\epsilon = T_{00}/\Lambda^4$ is depicted in Fig. 3. Fig. 4 presents the energy density for one of the Coulomb peaks from Fig. 3. In Fig. 5 the energy density profiles ($\epsilon(\rho)$ for $\phi = z = 0$) for various values of $R$ are compared. In all figures we use the rescaled spatial coordinates: $z\Lambda \to z$, $\rho\Lambda \to \rho$ and $R \to \Lambda R$.. It is clear that the flux-tube connecting the two point charges becomes thicker when the charge separation distance increases. We have obtained the same pictures starting from several different initial configurations for the iterative procedure. Our numerical procedure was also tested for $\delta = 0$. In this case the analitical solution for $\Phi$ is given by formula (36). The numerical procedure recovers it.

The energy density distribution was used to derive the total energy of the considered field configuration. As the energy density decreases slowly when $\rho^2 + z^2 \to \infty$ it is necessary to collect the energy density from relatively large region around the charges. For this reason we have used the lattice with link length varying from short near the charges to long for large $\rho^2 + z^2$. The energy of configurations for various $R$ was used to check the relation (26) - if it holds $\varepsilon$ should depend linearly on $\sqrt{R}$ for $\delta = 0.75$. Using numerical results for this linear relation one can estimate that $c_0 \approx 4.1$.

## 4 Summary and remarks

We have investigated the Abelian sector of Pagels-Tomboulis model. If the total charge $Q$ is non-vanishing the resulting field has infinite energy, while in the case of point-like charges $q, -q$ separated by the distance $R$ the energy is finite, proportional to $R^{(4\delta - 1)/(4\delta + 1)}$, provided that $\delta > 1/4$.

There are several directions in which one could continue the present work. First, we have not investigated stability of our Abelian solutions against fluctuations of gauge fields. Because the derivation of screening phenomenon given in the standard Yang-Mills case by Kiskis, [16], does not work if $\delta > 0$, one may hope that a presence of the other components of the gauge field $(A_\mu^1, A_\mu^2)$ will increase the energy. Also, one may ask about stability against a collapse of the field configuration in the Abelian sector. To answer this question one would have to investigate time-dependent solutions. It might turn out that additional terms would have to be included in the Lagrangian in order to prevent such a collapse, in an analogy to the well-known Skyrme term in a mesonic effective Lagrangian.

Finally, one could use more refined forms of the $\beta(g)$ function in Eq. (4). The model considered in the present paper should be regarded as the simplest one compatible with the asymptotic freedom.

All problems mentioned above are important, but in our opinion the most important and interesting one is to describe physical excitations in the models of the Pagels-Tomboulis type.

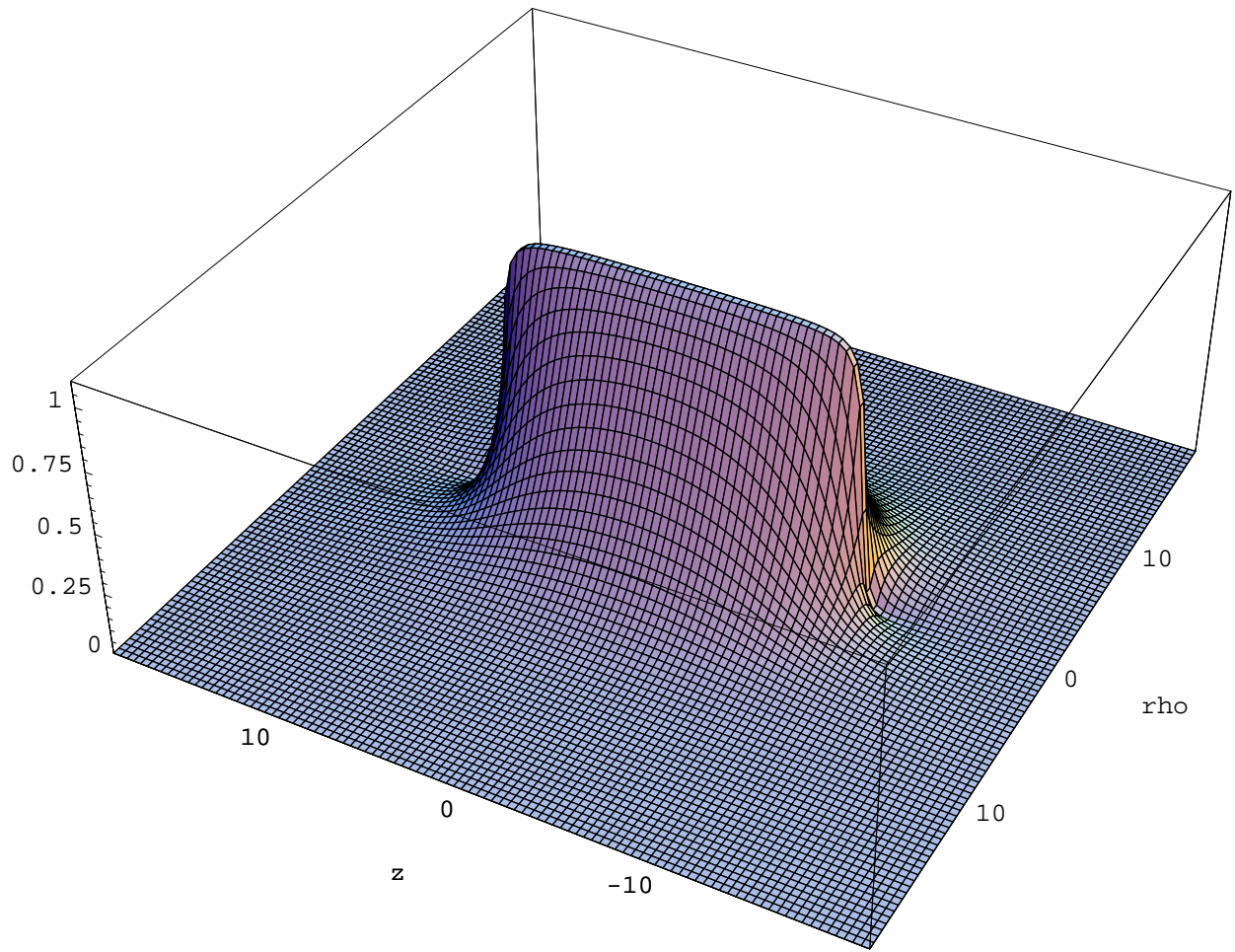

Figure 1: The flux function $\Phi$ for $\delta = 0.75$ and $R = 20$.

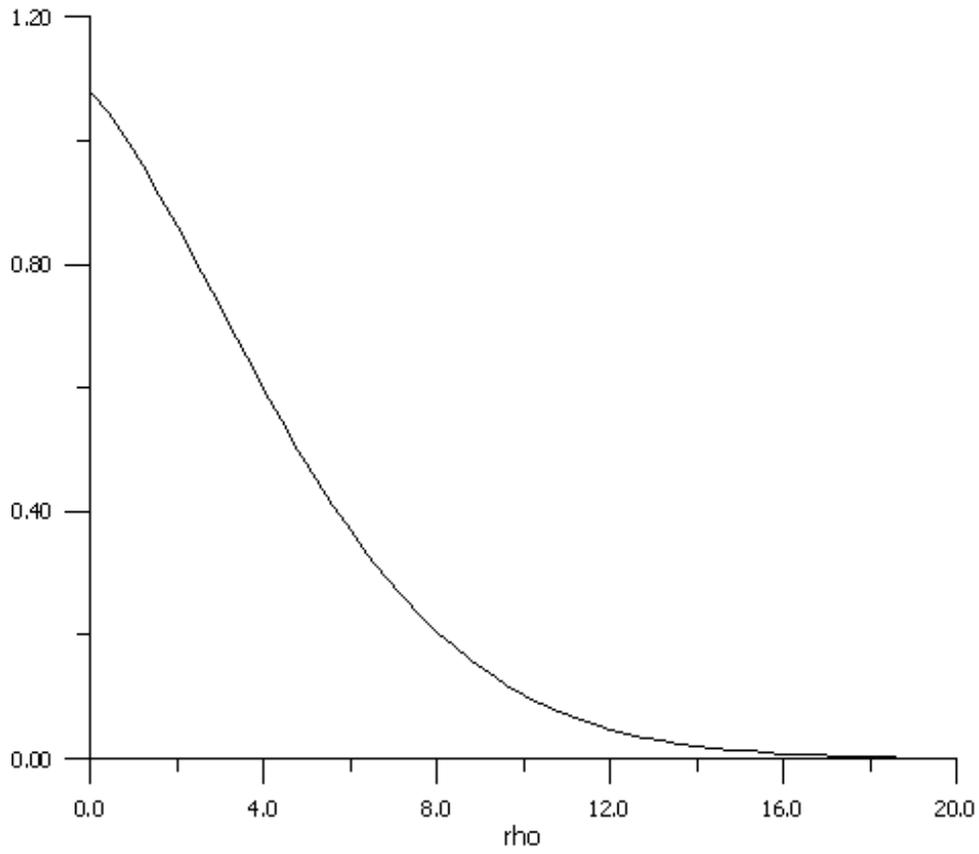

Figure 2: The flux function $\Phi$ for $\delta = 0.75$, $R = 20$ and $z = 0$.

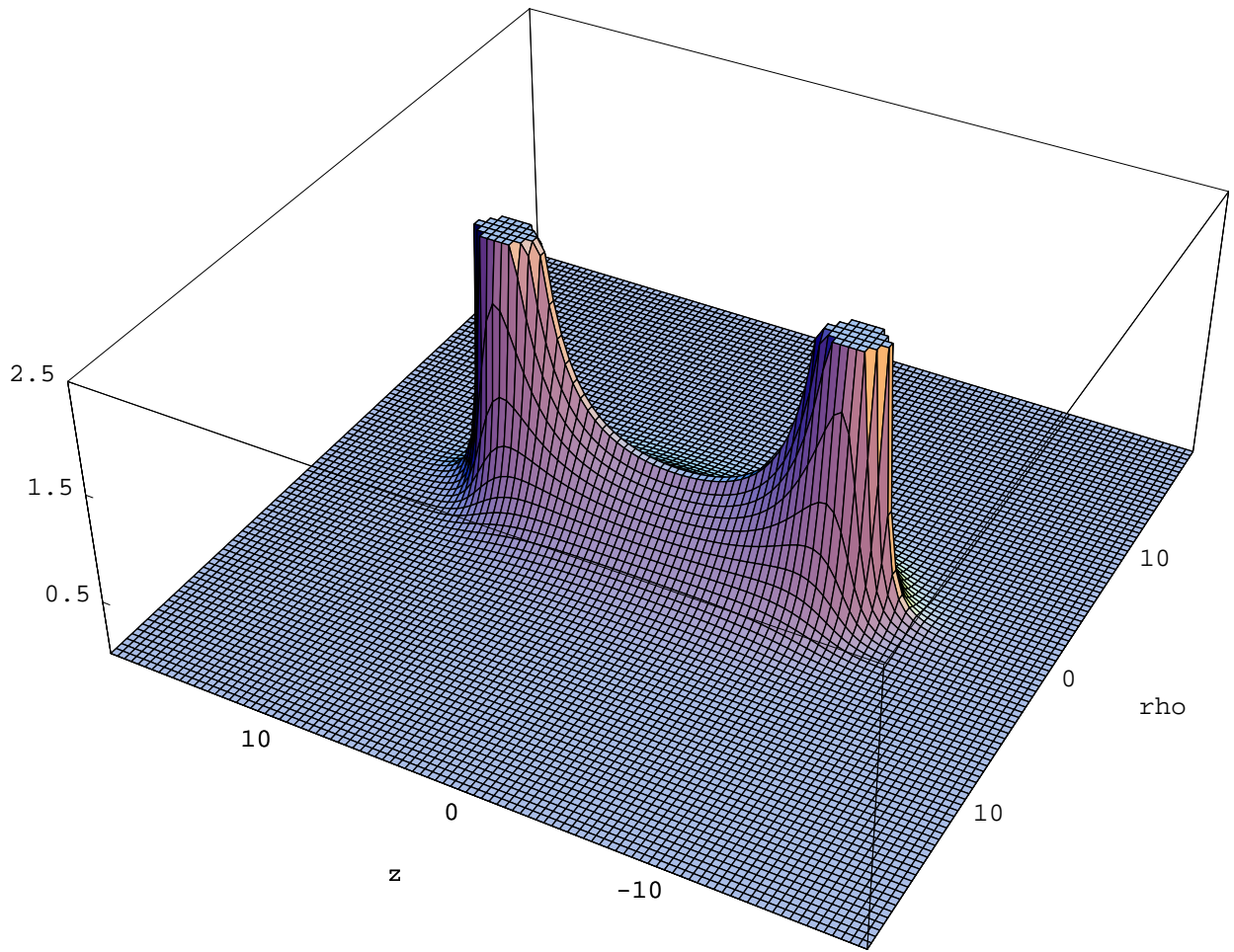

Figure 3: The energy density $\epsilon$ for $\delta = 0.75$ and $R = 20$.

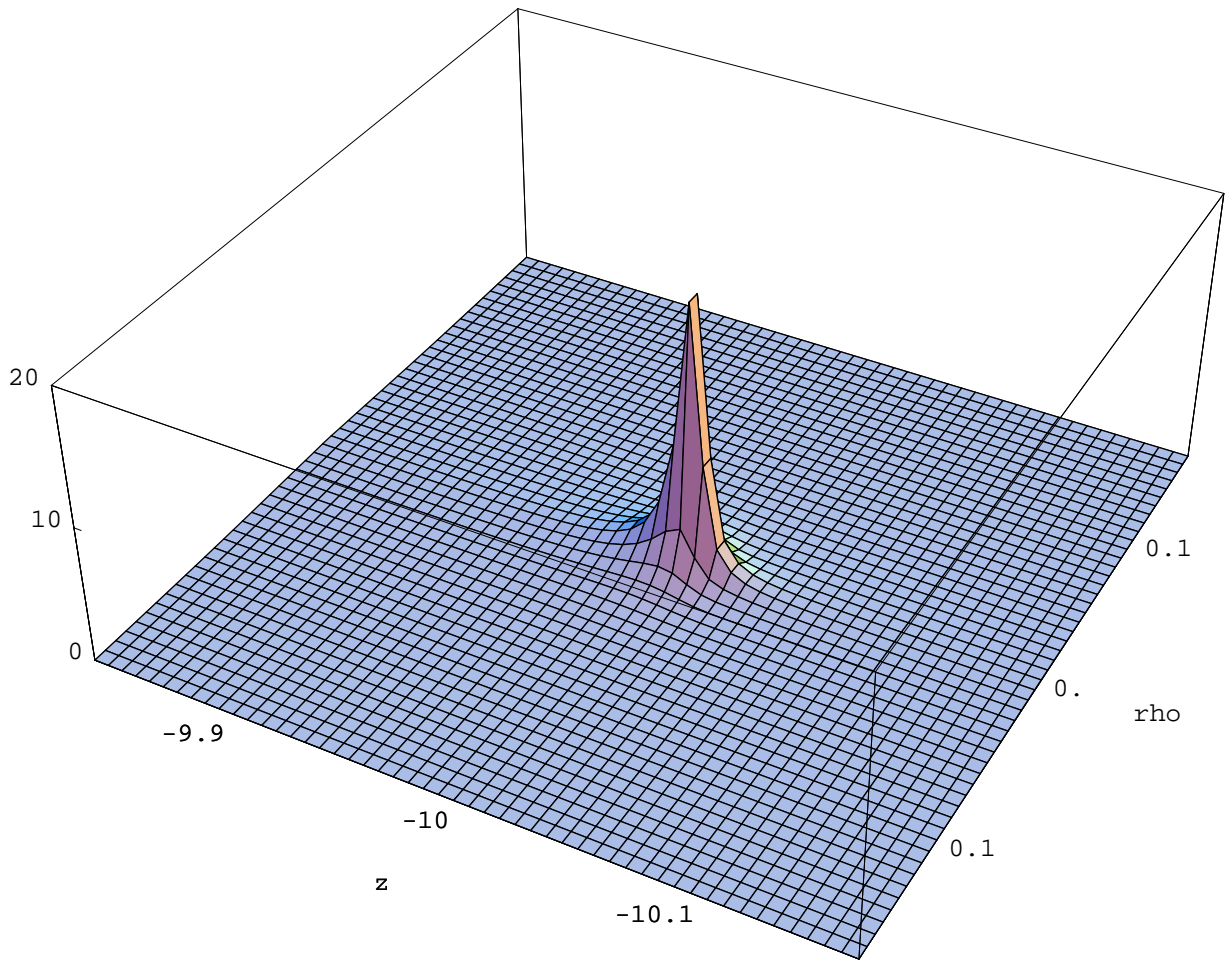

Figure 4: The energy density of the point charge $\epsilon$ for $\delta = 0.75$ and $R = 20$.

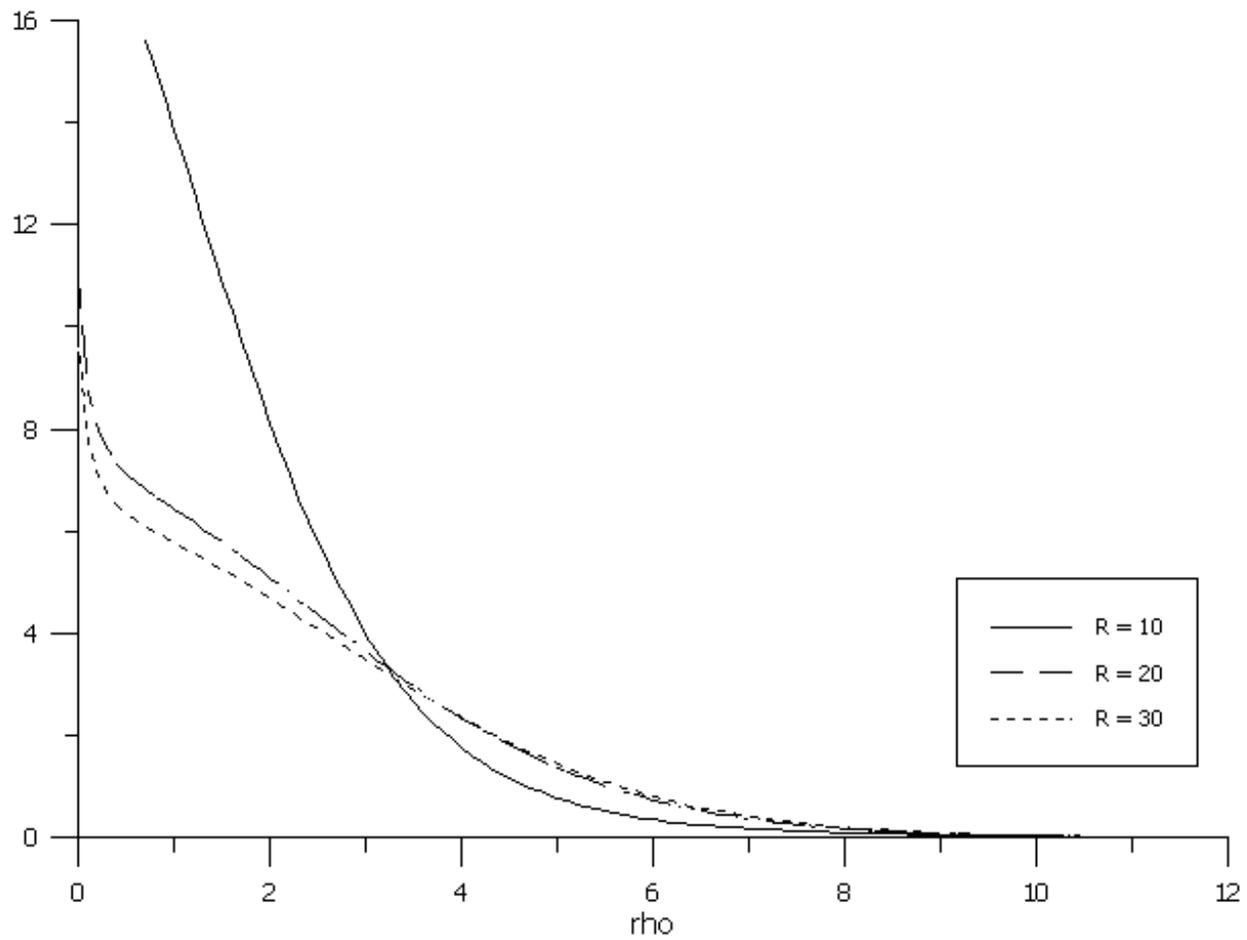

Figure 5: The energy density $\epsilon(\rho)$ for $\delta = 0.75$, $z = \phi = 0$ and $R = 10, 20, 30$.